\documentstyle[aps,prl,twocolumn]{revtex}
\begin{document}
\title{Universal saturation of
electron dephasing in three-dimensional disordered metals}
\author{J. J. Lin$^\ast$ and L. Y. Kao}
\address{Institute of Physics, National Chiao Tung University,
Hsinchu 300, Taiwan}
\date{\today}
\maketitle

\begin{abstract}
We have systematically investigated the low-temperature electron
dephasing times $\tau_\phi$ in more than 40 three-dimensional
polycrystalline impure metals with distinct material
characteristics. In all cases, a saturation of the dephasing time
is observed below about a (few) degree(s) Kelvin, depending on
samples. The value of the saturated dephasing time $\tau_0$
[$\equiv \tau_\phi (T \rightarrow 0\, {\rm K})$] falls basically
in the range 0.005 to 0.5 ns for all samples. Particularly, we
find that $\tau_0$ scales with the electron diffusion constant $D$
as $\tau_0 \sim D^{- \alpha}$, with $\alpha$ close to or slightly
larger than 1, for over two decades of $D$ from about 0.1 to 10
cm$^2$/s. Our observation suggests that the saturation behavior of
$\tau_\phi$ is universal and intrinsic in three-dimensional
polycrystalline impure metals. A complete theoretical explanation
is not yet available.

\end{abstract}

The issue concerning the saturation of conduction electron
dephasing time $\tau_\phi(T)$ in disordered conductors and
mesoscopic systems has recently attracted renewed theoretical
\cite{Zaikin98,Alt98,Zawa99,Imry99} and experimental
\cite{Webb97,Gershen98,Bird99,Marcus99} attention. In particular,
it is of great interest in whether or not the saturation of
$\tau_\phi$ might be universal in all material (e.g.,
polycrystalline metal, amorphous metal, and MBE-grown
semiconductor) systems and in all (zero, one, two, and three)
dimensions. It is also of importance to ask whether or not there
might be a common, intrinsic source that causes the (almost)
universally observed saturation. The interest in this issue of the
existence or not of a {\em finite} conduction electron dephasing
time at absolute zero temperature is closely connected with the
concern about the validity of the Fermi-liquid behavior, the
possibility of the occurrence of a quantum phase transition, and
also the feasibility of quantum computing \cite{Mohanty00}. There
are also works suggesting that this might explain the
long-standing persistent current problem in metals \cite{persist}.
In practice, the value of $\tau_\phi$ can be reliably determined
using quantum-interference studies
\cite{Webb97,Gershen98,Bird99,Marcus99} such as weak-localization
effects and universal conductance fluctuations. It can also be
determined using studies of the shape change of the electron
distribution function upon applying a bias voltage across the
conductor \cite{Birge00}.

In this work, we have systematically investigated the electron
dephasing times $\tau_\phi$ in numerous {\em three}-dimensional
{\em polycrystalline} impure metals. Our samples were made of
various materials using various fabrication techniques at
different laboratories. Also, the samples were measured at
different laboratories at very different times. Our samples
considered in this work include dc sputtered thick
Au$_{50}$Pd$_{50}$ films, dc and/or RF sputtered thick
Ag$_{40}$Pd$_{60}$ films, dc sputtered thick Sb films,
thermal-evaporation deposited thick Au$_x$Al films ($1.8 \lesssim
x \lesssim 2.2$), thermal-evaporation deposited thick
Sc$_{85}$Ag$_{15}$ films, and arc-melted V$_{100-x}$Al$_x$ alloys
($20 \lesssim x \lesssim 24$). The thick film samples are
typically on the order 4000 $\rm\AA$ $\times$ 0.3 mm $\times$ 17
mm, while the arc-melted samples are typically on the order 0.1
$\times$ 0.1 $\times$ 10 mm$^3$. Since more than 40 samples are
studied, it would not be practical to list the material parameters
for all the samples. However, we notice that all of the samples
studied are made from very high-purity starting materials obtained
from reputable suppliers such as Alfa Aesar, Cerac, and
Goodfellow. Since the major aim of this work is to study
experimentally whether there might exist a universal behavior of
the saturated (or, the zero-temperature) dephasing time $\tau_0$
[$\equiv \tau_\phi (T \rightarrow 0\, {\rm K})$], the use of many
kinds of samples with very distinct characteristics should
therefore serve well this purpose. Any behavior of $\tau_0$ common
for all these materials, if observed, must be a manifestation of
the very general nature of the zero-temperature dephasing time. In
addition, we notice two more practices regarding our experimental
method. (a) Our measurements of $\tau_\phi$ had been performed
over a long time period of over three years (1997-2000). (b) Our
measurements of $\tau_\phi$ were carried out at two different
laboratories located at two different sites 80 km apart.
Nevertheless, the same $^3$He fridge and the same electronic
measuring systems (except with different grounding) were used for
all measurements. Surprisingly, regardless of all the
above-mentioned very different preparation and measurement
conditions, it is noteworthy that our $\tau_0$ measured in all
samples varies with the electron diffusion constant $D$ with a
simple power law as $\tau_0 \sim D^{- \alpha}$, with $\alpha$
close to or slightly larger than 1 (see below). Such universal
behavior could not be just ``accidental" and certainly deserves
serious theoretical and experimental attention. We emphasize that
we are concerned with three-dimensional polycrystalline metals. It
is our opinion that the saturation behavior (e.g., the functional
form of $\tau_0$ on disorder) is universal for a given
dimensionality and a given kind of sample (e.g., polycrystalline
or well-textured semiconductor) structure, while it might not be
universal over different dimensionalities and different sample
structures.

We have measured the low-field magnetoresistances at liquid-helium
temperatures and compared with three-dimensional weak-localization
predictions to extract the values of $\tau_\phi(T)$ for our
samples. The details of the data analysis procedure had been
discussed previously \cite{Lin94}. Here we merely stress that, for
every sample studied in this work, the three-dimensional
weak-localization predictions can well describe our experimental
measurements. Thus, $\tau_\phi$ can be very reliably extracted.
Figure~\ref{fig1} shows the total electron dephasing rate
$\tau_\phi^{-1}$ as a function of temperature for three
representative thick Ag$_{40}$Pd$_{60}$ films with resistivities
as indicated in the caption to Fig. \ref{fig1}. The symbols are
experimental data and the solid curves (emphasizing particularly
the low measurement temperatures) are the least-squares fits to
the form
\begin{equation}
\tau_\phi^{-1}(T) = \tau_0^{-1} + \tau_{\rm i}^{-1}(T) \,\,,
\end{equation}
where we assume that the total dephasing time comprises two terms,
with $\tau_0$ being a temperature-independent (i.e., the
saturated) dephasing time which dominates at the lowest
measurement temperatures, and $\tau_{\rm i}$ being the relevant
inelastic scattering time which is usually dominant at a (few)
degree(s) Kelvin or so and higher, strongly depending on samples.
One sees that, for every sample, $\tau_\phi^{-1}$ first decreases
with decreasing temperature in the high-temperature part of our
measurements. This decrease in the total dephasing rate can
readily be ascribed to the weakening of the inelastic scattering
process with decreasing temperature. That is, $\tau_\phi^{-1}$ is
dominated by $\tau_{\rm i}^{-1}$ and it decreases with weakening
$\tau_{\rm i}^{-1}$ above 3 K or so for these particular samples.
At lower temperatures, the inelastic scattering becomes weaker and
weaker so that $\tau_\phi^{-1}$ becomes more and more dominated by
the temperature-independent $\tau_0^{-1}$. Inspection of Fig.
\ref{fig1} reveals that Eq. (1) can well describe the experimental
data. In fact, in some cases (e.g., Fig. \ref{fig1}) when the
saturation of $\tau_\phi$ already sets in at a few degrees Kelvin
high, $\tau_0$ can already be determined using essentially the
measured value at the lowest temperatures. This ``advantage" of
(relatively) {\em high}-temperature saturation largely reduces the
uncertainties in the least-squares fits, making the inferred
values of the adjusting parameters $\tau_0$ and $\tau_{\rm i}$
very reliable.

Before the recent renewed interest in $\tau_0$, a saturation of
the electron dephasing time as that shown in Fig. \ref{fig1} has
been observed in a good number of experiments, including
reduced-dimensional metals \cite{Lin87} and semiconductors
\cite{Pepper89}. One of the most adopted early explanations
proposed for the observed saturation invokes hot-electron effects.
However, this kind of explanation can be ruled out for our
measurements in Fig. \ref{fig1}. In the hot-electron picture, it
is argued that if the Joule heating caused by measurement currents
were sufficiently large, the conduction electrons would be removed
out of thermal equilibrium with the phonon bath. Then, the
electron temperature would be higher than the phonon (lattice)
temperature, resulting in a ``roll-over" of $\tau_\phi$ at the
lowest measurement temperatures. Since we are concerned with
three-dimensional systems in this work, Joule heating is {\em
negligible} in our case. To confirm this assertion experimentally,
we plot in Fig. \ref{fig2} the variations of resistivity with
temperature for those representative thick Ag$_{40}$Pd$_{60}$
films just considered in Fig. \ref{fig1}. Comparison examination
of Figs. \ref{fig1} and \ref{fig2} indicates that the
resistivities of the samples keep increasing with reducing
temperature all the way down to our lowest measurement
temperatures (400 mK), though a saturation in $\tau_\phi$ is
already seen at a much higher temperature ($\sim$ 3 K in these
particular samples). Indeed, the resistivity rises with decreasing
temperature given in Fig. \ref{fig2} can essentially be described
by disorder enhanced electron-electron interaction effects
\cite{Alt85}, suggesting that the temperature of the electron gas
is the same to the temperature of the lattice. Therefore, {\em no}
appreciable Joule heating was produced in our measurements.
Consequently, the saturation of $\tau_\phi$ shown in Fig.
\ref{fig1} can {\em not} be interpreted in terms of hot-electron
effects. This is true for all samples studied.

The central result of this work is plotted in Fig. \ref{fig3}.
Figure~\ref{fig3} shows a plot of our measured saturated dephasing
time $\tau_0$ as a function of the electron diffusion constant $D$
for more than 40 samples made from very different materials and
prepared with very different fabrication techniques. Different
symbols label different material systems as indicated in the
caption to Fig. \ref{fig3}. In all cases, the samples are
three-dimensional with regard to the electron phase coherence
length $\sqrt{D \tau_\phi}$. Inspection of Fig. \ref{fig3} reveals
that the saturated dephasing time $\tau_0$ for all samples falls
essentially on the same region in the coordinates. In particular,
it reveals that all that matter in determining the value of
$\tau_0$ is the diffusion constant $D$, regardless of the distinct
material characteristics (e.g., electronic structures) of the
various samples. (It is understood that the evaluation of the
diffusion constant $D$ for some metal alloys could probably be
subject to an uncertainty of, e.g., as large as a factor $\sim$ 2,
due to the complex electronic structures. However, our
experimental availability of a very wide range of $D$ ensures that
even with a factor of 2 in the uncertainty of $D$ would not change
the main conclusion of this work \cite{diffusion}.) Close
inspection of Fig. \ref{fig3} indicates that $\tau_0$ varies
essentially with the diffusion constant $D$ as $\tau_0 \sim D^{-
\alpha}$, with $\alpha$ close to or slightly larger than 1, for
about {\em two decades} of $D$. As $D$ increases from about 0.1 to
10 cm$^2$/s (corresponding to the product $k_Fl$ ranging from of
order unity to of order 10, where $k_F$ is the Fermi wave number,
and $l$ is the electron elastic mean free path), $\tau_0$
decreases correspondingly from about 0.5 to 0.005 ns. In this
figure, the straight solid line is drawn proportional to $D^{-1}$
and is a guide to the eye. This observed $\tau_0 \sim D^{-
\alpha}$, with $\alpha$ close to or slightly larger than 1, is
totally unexpected. On the contrary, it is often conjectured that
$\tau_0$ should increase with {\em reducing} disorder, {\em at
least} in one and two dimensions \cite{Webb97}. Until now, it is
{\em not} known exactly how differently the saturated dephasing
time should behave in different dimensionalities and in different
sample structures. Surprisingly, we notice that our observation of
an essentially inverse linear dependence of $\tau_0$ on $D$
actually implies an essentially ``constant" saturated dephasing
length of $\sqrt{D\tau_0} \sim$ 1000 $\rm\AA$ in all samples. For
comparison, the experimental values of $\tau_0$ previously
reported by Lin and Giordano \cite{Lin87} in thin, polycrystalline
AuPd films are also indicated in Fig. \ref{fig3} by the two
vertical bars: the vertical bar at $D \approx$ 2.2 cm$^2$/s stands
for the range of $\tau_0$ observed in their dc sputtered films,
while the vertical bar at $D \approx$ 23 cm$^2$/s stands for that
in their thermally evaporated films. The observation of Fig.
\ref{fig3} is for the first time such a systematic study has been
done on so many samples with a wide range of diffusion constant.

One of the most popular (early) explanations for the saturation of
$\tau_\phi$ invokes spin-spin scattering due to the presence of a
very minor amount of magnetic impurities in the sample. This
explanation was refuted in many experiments \cite{Webb97,Lin87}.
In the present work, this explanation can also be ruled out simply
because our samples were made from very different high-purity
sources obtained from various reputable suppliers, and measured at
very different times. It is hard to conceive that magnetic
spin-spin scattering due to ``accidental" contamination could have
caused the ``systematic" variation of $\tau_0$ with $D$ as
observed in Fig. \ref{fig3}. Moreover, since we are concerned with
bulk samples, any spin-flip scattering that might result from
surface effects, such as interfaces, substrates, and paramagnetic
surface oxidation, can largely be minimized (while surface effects
could be significantly more important in reduced-dimensional
systems). Indeed, as mentioned, we do {\em not} see a logarithmic
Kondo type of dependence in resistivity (Fig. \ref{fig2}) in our
temperature range, strongly ruling out magnetic impurities playing
any role in our samples. It is also worthwhile pointing out in
passing that, although it has been widely taken for granted for
over years that magnetic impurities will cause dephasing and
produce a temperature independent $\tau_\phi$, Webb and coworkers
\cite{Webb97} have recently ruled out magnetic impurities as the
cause of the observed saturation.

The origin for the observed saturated dephasing time given in Fig.
\ref{fig3} is not exactly clear. To our knowledge, Altshuler,
Aronov, and Khmelnitsky \cite{Alt98,Alt81} have considered the
dephasing of electron wave amplitudes by high-frequency
electromagnetic noises. They found that an electromagnetic noise
can be, on the one hand, already large enough to cause dephasing
while, on the other hand, still too small to cause significant
Joule heating of the conduction electrons. Unfortunately, in the
most effective frequency range in causing dephasing, their theory
predicts a form $\tau_0 \sim D^{-1/3}$ which is in disagreement
with our experimental observation. Besides, recent experiments
\cite{Marcus99,Webb98} has explicitly demonstrated that direct
dephasing due to radiation could not be the cause of the observed
saturation. Therefore, we have to turn to other explanations, such
as, among others, zero-point fluctuations of the electrons
\cite{Zaikin98}, dynamical two level systems \cite{Zawa99}, and
coherent charge transfer between crystallites \cite{GYWu00}.
Unfortunately, in terms of the zero-point fluctuations theory, we
observe, in almost all samples at 500 mK, a $\tau_0$ already
exceeds 3 to 5 orders in magnitude above the upper bound set by
the theory. [The saturation time in three dimensions is given by
the zero-point fluctuations theory as $\tau_{\rm GZ}^{\rm (3D)} =
(22\pi^2 \hbar /e^2)(D^2/\rho_0 v_F^3)$, where $\rho_0$ is the
residual impurity, and $v_F$ is the Fermi velocity. See, e.g.,
Refs. \cite{Zaikin98,Alt98}.] Even worse, the zero-point
fluctuations theory predicts increasing $\tau_0$ with increasing
$D$, opposite to our observation in {\em polycrystalline} impure
metals. Although an increase of $\tau_0$ with increasing mobility
has been observed in, e.g., {\em semiconductor quantum wires}
\cite{Ikoma96} and {\em quantum dots} \cite{Bird99,Marcus99}, it
is conceived \cite{GYWu00} that the saturation behavior in
polycrystalline impure metals could be different from that in
well-textured semiconductor quantum wires and dots. Indeed, it is
our opinion that the saturation behavior should be universal among
polycrystalline impure metals, while different functional forms of
$\tau_0$ on disorder might be expected for different sample
structures. If this were the case, then the current concept for
zero-temperature dephasing would need major redirection.

In conclusion, we have observed a saturated dephasing time
$\tau_0$ on the order of 0.005 to 0.5 ns in more than 40
three-dimensional polycrystalline impure metals with distinct
characteristics. Taken many metals and alloys together, our result
indicates that $\tau_0$ varies essentially with the electron
diffusion constant $D$ as $\tau_0 \sim D^{- \alpha}$, with
$\alpha$ close to or slightly larger than 1, for about two decades
of $D$. It is for the first time such a systematic study has been
done on so many samples with a wide range of diffusion constant. A
complete theoretical explanation for this observation is not yet
available. On the other hand, our result suggests that the origin
for the observed saturation at low temperatures might not be
universal over all dimensionalities and all kinds of sample
structures. If this were the case, then the current concept for
zero-temperature dephasing would need major redirection.

The authors are grateful to T. J. Li, S. B. Chiou, S. F. Chang, Y.
L. Zhong, and A. K. Meikap for help in the experiment. This work
was supported by Taiwan National Science Council through Grant No.
NSC 89-2112-M-009-033.

\begin{figure}
\caption{Electron dephasing rate $\tau_\phi^{-1}$ as a function of
temperature for three thick Ag$_{40}$Pd$_{60}$ films having
$\rho$(10\,K) = 119 ($\Box$), 178 ($\circ$), and 281 ($\triangle$)
$\mu \Omega$ cm. The solid curves are least-squares fits to Eq.
(1). \label{fig1}}

\vspace{1cm} \caption{Normalized resistivity $\triangle \rho
(T)$/$\rho$(10\,K) = [$\rho(T) - \rho$(10\,K)]/$\rho$(10\,K) as a
function of $\sqrt{T}$ for three thick Ag$_{40}$Pd$_{60}$ films
having $\rho$(10\,K) = 119 ($\Box$), 178 ($\circ$), and 281
($\times$) $\mu \Omega$ cm. \label{fig2}}

\vspace{1cm} \caption{Variation of saturated dephasing time
$\tau_0$ with electron diffusion constant $D$ for various
three-dimensional polycrystalline impure metals:
Au$_{50}$Pd$_{50}$ (circles), Ag$_{40}$Pd$_{60}$ (squares), Sb
(triangles), Au$_x$Al (solid triangles), Sc$_{85}$Ag$_{15}$ (solid
squares), and V$_{100-x}$Al$_x$ (solid circles). The two vertical
bars at $D \approx$ 2.2 and 23 cm$^2$/s represent the experimental
values of $\tau_0$ for dc sputtered and thermally evaporated thin,
polycrystalline AuPd films, respectively, taken from Ref. [13].
The straight solid line is drawn proportional to $D^{-1}$ and is a
guide to the eye. \label{fig3}}

\end{figure}

\begin{references}
\bibitem{diffusion} The values of $D$ for most of our samples were
determined through the Einstein relation $D = [\rho_0 e^2
N(0)]^{-1}$, using the measured values for the impurity
resistivity $\rho_0$, and the values of electronic density of
states at the Fermi level $N(0)$ were evaluated through measured
low-temperature specific heats.
\end{references}
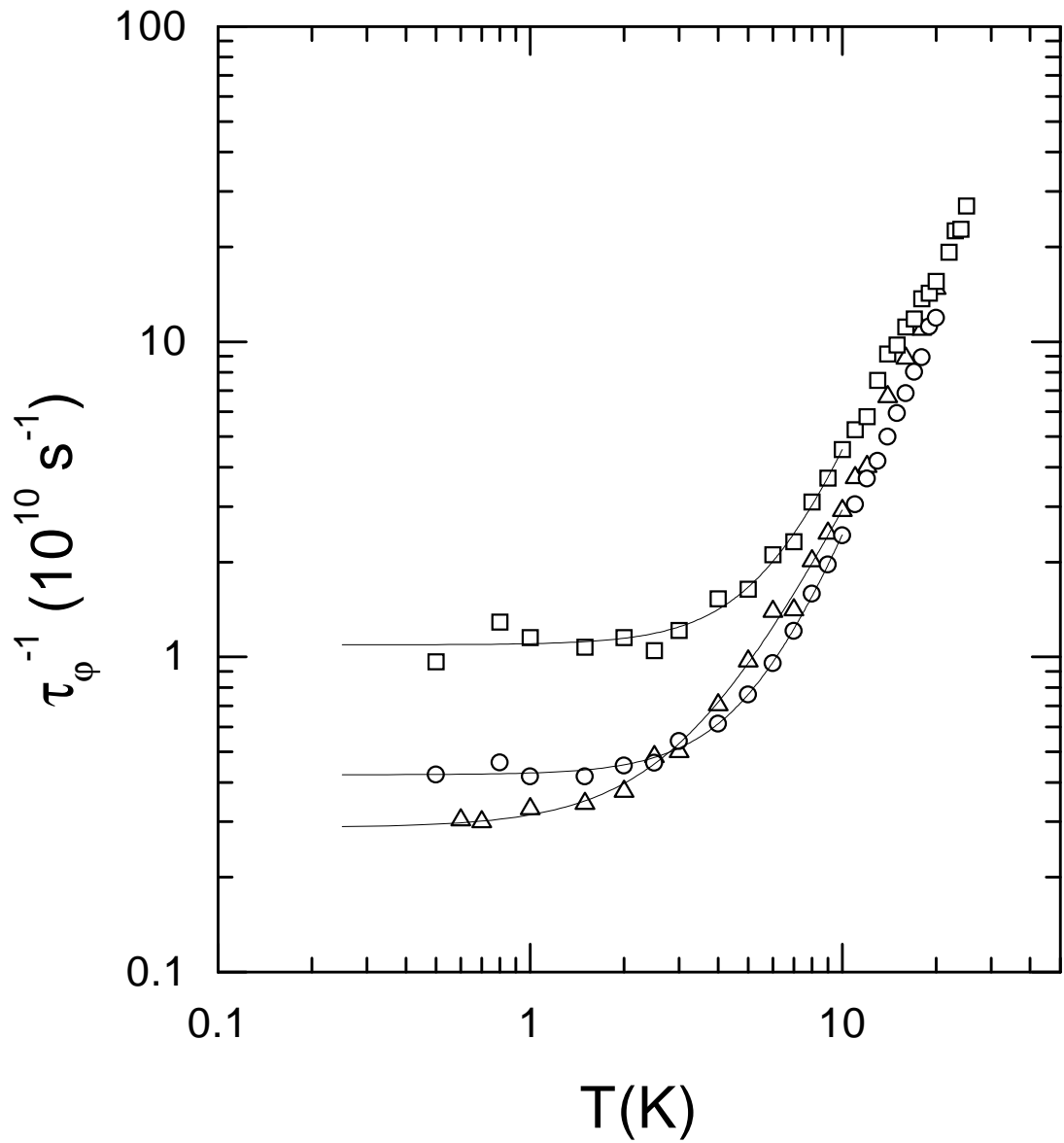
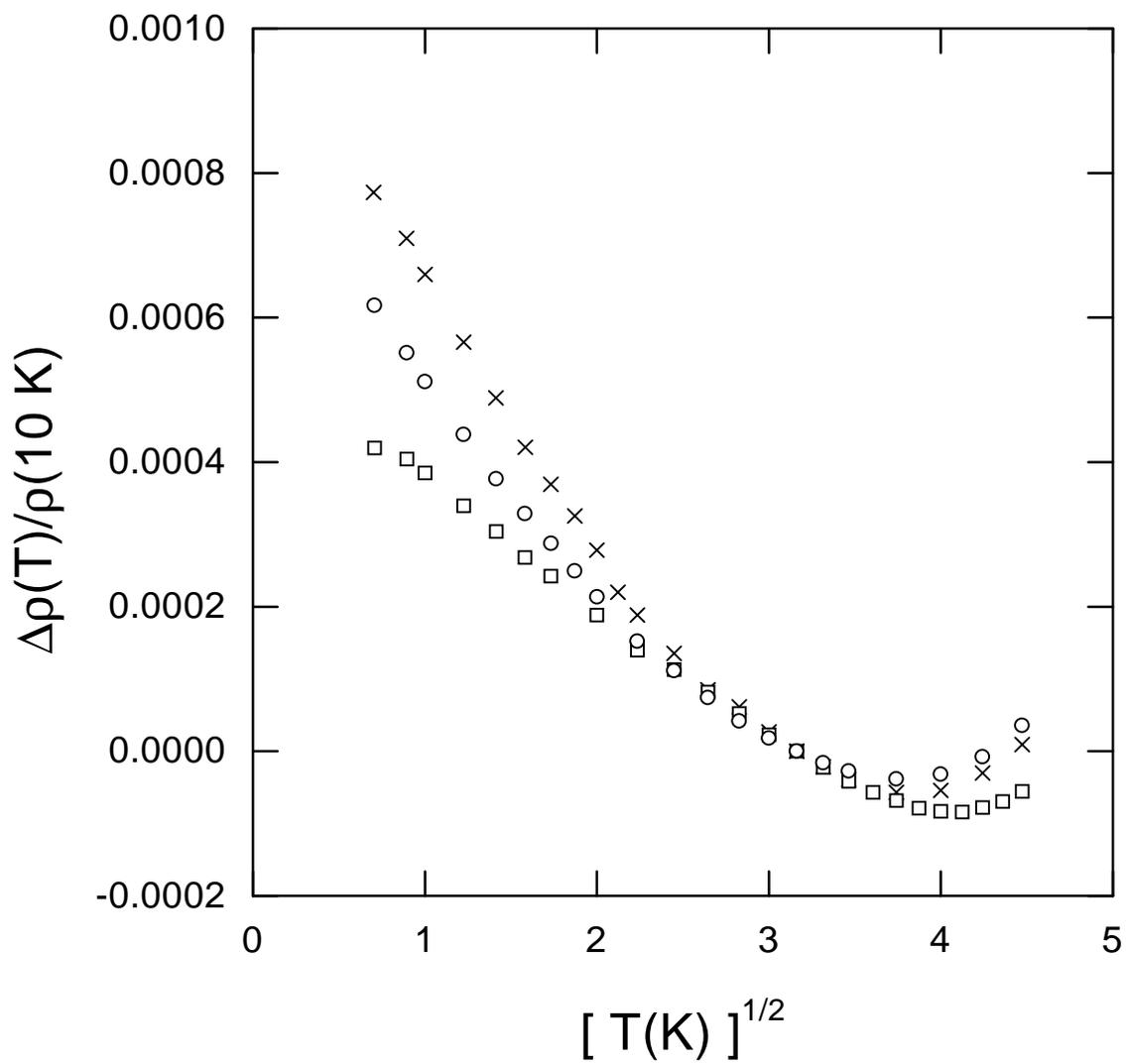
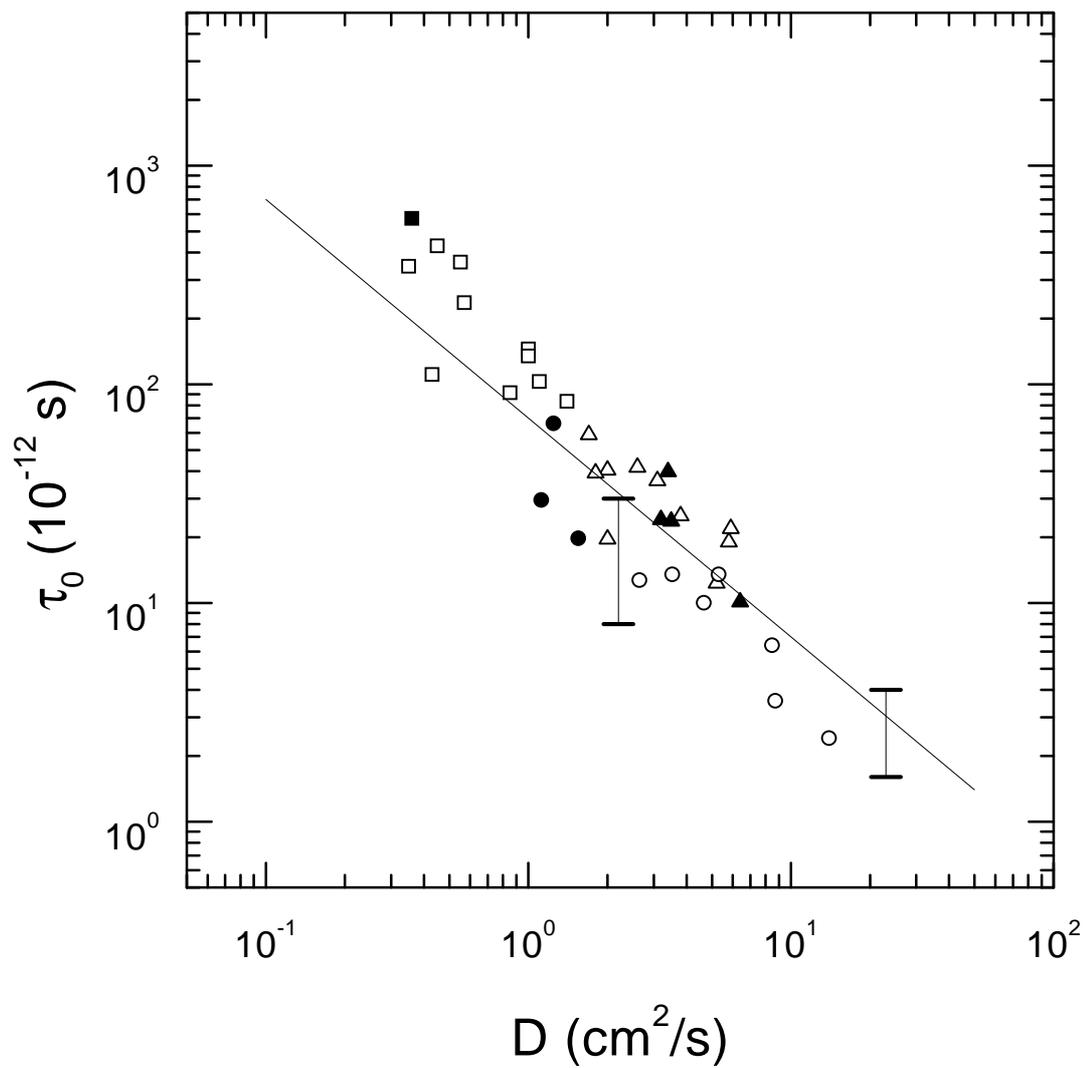
\end{document}